# Mutually Unbiased Bases and Orthogonal Latin Squares

Stefan Joka

*Faculty of Physics, University of Belgrade, Studentski Trg 12-16, 11000 Belgrade, Serbia*

*e-mail: stefanjoka35@gmail.com*

**Abstract**

In this paper, we prove that the existence of a complete set of mutually unbiased bases (MUBs) in $N$- dimensional Hilbert space implies the existence of a complete set of mutually orthogonal Latin squares (MOLSs) of order $N$. In particular, we prove that a complete set of MUBs does not exist in dimension six (the first dimension which is not a power of prime).

## Introduction

Two orthonormal bases $\{ | f_i \rangle \}$ and $\{ | g_k \rangle \}$ in $N$- dimensional Hilbert space $\boldsymbol{C}^N$ are said to be mutually unbiased if, for all $i, k$ :

$$|\langle f_i | g_k \rangle|^2 = \frac{1}{N},$$

where $\langle f_i | g_k \rangle$ is a standard scalar product of vectors $|f_i\rangle$ and $|g_k\rangle$.

If we look at the orthonormal bases as a sets of orthogonal projectors belonging to the space of $N \times N$ complex matrices $M_N(\boldsymbol{C})$, the above formula becomes:

$$\mathrm{Tr}(P_i P_k) = \frac{1}{N},$$

where $P_i$ and $P_k$ are the rank-one projectors from two mutually unbiased bases.

In quantum theory, MUBs express the principle of complementarity: two observables corresponding to a pair of MUBs cannot be known simultaneously.

About the number of MUBs, the following is known:

$a$) In any dimension $N$, the maximal number of MUBs is $N + 1$ [2]; the set of $N + 1$ MUBs is called complete;

$b$) In any prime-power dimension, there exists a complete set of MUBs [2];

$c$) In any dimension $N$, the set of $N$ MUBs can be completed [3];

$d$) Little is known about the maximal number of MUBs in dimensions that are not prime powers; the question is still open even for dimension six.

It has been realized that the MUB problem is closely related to some other problems in mathematics and physics. Here, we mention an open problem of describing orthogonal pairs and orthogonal decompositions in Lie algebra theory [10] as well as a combinatorial problem of determining the maximal number of mutually orthogonal Latin squares (MOLSs).

Problem of MOLSs, initiated by Euler, is still unsolved, although it is known more about it than about the problem of MUBs. For example, it is known that a complete set of MOLSs does not exist for $N = 6$ and $10$ [1]. It is also known that a complete set of MOLSs does exist in prime-power dimensions and a formal equivalence between two problems (MUBs and MOLSs), for $N$ being a power of prime, has been established [4].

In this paper, we prove that, for any $N$, the existence of a complete set of MUBs implies the existence of a complete set of MOLSs. To prove that, we combine the known results stemming from different approaches to the MUB problem (complementarity polytope, maximal abelian subalgebras or maximal commuting basis) and some results from the theory of symplectic toric manifolds.

The paper is organized as follows: in sections 1 and 2, we state the known results about the complementarity polytope and a decomposition of $M_N(\boldsymbol{C})$ into maximal abelian subalgebras and their link to MUBs. In section 3, we recall some of the results from the theory of symplectic toric manifolds which will be an important part of our proof. In section 4 we give the proof of aforementioned implication.

## 1. Complementarity polytope

In this section, we explain the idea of complementarity polytope which was studied in detail by Bengtsson and Ericsson in [1]. All results listed here are known and can be found in [1].

The idea of Bengtsson and Ericsson is to look at MUBs as the elements of the space of density matrices, that is, the subset of the space of Hermitian matrices of unit trace. This is an $(N^2-1)$ real dimensional space. Basis vectors are represented by the rank-one projectors and the distance between two points is given by :

$$d^2(A,B) \;= \frac{1}{2}\mathrm{Tr}(A-B)^2.$$

From the above formula it is obvious that the orthogonal projectors from one orthonormal basis form a regular $(N-1)-$ simplex (we will call it a $P$-simplex) since all of them sit at the same distance from each other. If we choose the matrix $\frac{1}{N}I$ ($I$ is the unit matrix) as the origin of our space, scalar product becomes:

$$\langle A|B\rangle = \frac{1}{2}\left[\mathrm{Tr}(AB) - \frac{1}{N}\right].$$

Therefore, we see that the $P$-simplices, corresponding to mutually unbiased bases, are orthogonal to each other and since the dimension of the whole space is $N^2-1$, the maximal number of MUBs is $N+1$. Now, suppose that this $(N+1)$ bases exist. We can construct a convex hull of their $P$-simplices and obtain a polytope, called the complementarity polytope, in the $(N^2-1)$ real dimensional space. Actually, the complementarity polytope can be constructed in any dimension, but this does not imply the existence of a complete set of MUBs, because one cannot be sure whether the vertices of the polytope would correspond to a pure states (the dimension of the outsphere of the polytope is $N^2-2$, while the set of pure states forms a $2(N-1)$-dimensional submanifold on the outsphere).

So, the mere existence of the complementarity polytope does not say much about the existence of a complete set of MUBs. In [1] Bengtsson and Ericsson came up with an idea to examine the possibility of inscribing a regular simplex (we will call this simplex an $A$ simplex) of dimension $N^2-1$ into complementarity polytope such that $N^2$ vertices of the simplex sit at the centers of $N^2$ facets of the polytope. **They proved that this is possible if and only if there exists a complete set of**

**MOLSs of order $N$**. We will use this result in our proof. Actually, we will see that both complementarity polytope and $A$ simplex will naturally arise from a symplectic geometry of a suitably chosen complex projective space and its subspaces.

## 2. Maximal commuting basis and quasi-orthogonal, maximal abelian subalgebras

In this section, we move on to study an algebraic approach to the MUB problem. All results listed here are known and can be found in [2] and [3].

We consider the space of $N \times N$ complex matrices $M_N(\boldsymbol{C})$ with scalar product $\langle A|B\rangle = \mathrm{Tr}\left(A^{\dagger}B\right)$. Now, one orthonormal basis $B_j$ in $\boldsymbol{C}^N$ consists of $N$ orthogonal rank-one projectors $P_k^j$, where index $j$ denotes a basis and index $k$ denotes a vector in that basis. Suppose, now, that two bases $B_i$ and $B_j$ are mutually unbiased. Then, we have the following relations for scalar products:

$$\left\langle P_s^j \middle| P_t^j \right\rangle = \delta_{s,t};$$

$$\left\langle P_s^i \middle| P_t^j \right\rangle = \frac{1}{N}.$$

Now, we state the theorem proved by Bandyopadhyay et al in [2] (this is the **theorem 3.4** in [2]).

**Theorem (Bandyopadhyay et al , [2]).** Let $B_1, \ldots, B_m$ be a set of MUBs in $\boldsymbol{C}^N$. Then there are $m$ classes $\varepsilon_1, \ldots, \varepsilon_m$ each consisting of $N$ commuting unitary matrices such that matrices in $\varepsilon_1 \cup \ldots \cup \varepsilon_m$ are pairwise orthogonal.

For a full proof see [2]. Here, we describe how to construct aforementioned classes. Each commuting class $\varepsilon_j$ is constructed from one of the MUBs ie. from projectors $P_k^j$ with $j$ fixed:

$$\varepsilon_j = \left(A_{j,0}, A_{j,1}, \ldots A_{j,N-1}\right),$$

where $A_{j,t} = \sum_{k=1}^{N} e^{2\pi i \frac{tk}{N}} P_k^j, \; 0 < t < N-1, \; 1 < j < m.$

Operators $A_{j,t}$ (from now on, we will call them $A$ -operators) are, by construction, unitary and, with exception of $A_{j,0} = I$, of vanishing trace. $A$-operators from the same class, are, also by construction, commuting and pairwise orthogonal. **The key point is: when two bases $B_i$ and $B_j$ are mutually unbiased, than $A$ -operators from corresponding classes are also pairwise orthogonal** (with exception of identity matrix $I$ not being orthogonal to itself). All these statements can be proved by using an elementary linear algebra (see [2]) Now, if we have a complete set of MUBs, there exist $N+1$ such commuting classes or $N^2$ orthogonal $A$ -operators which constitute one orthogonal basis (we will call it an $A$ –basis) for $M_N(\boldsymbol{C})$, basis called maximal commuting basis in [2].

If we look at the linear span of projectors from one MUB, we will see that $\mathrm{Span}\left(P_1^j, \dots, P_N^j\right) = \mathrm{Span}\left(A_{j,0} \dots A_{j,N-1}\right)$ is a subspace of $M_N(\boldsymbol{C})$ and moreover, it is a maximal abelian subalgebra of $M_N(\boldsymbol{C})$[3]. An approach to the MUB problem, via maximal abelian subalgebras (MASAs), is very well explained by Weiner in [3]. Now, theorem 3.4 from [2], in the case of complete set of MUBs, can be rephrased in terms of MASAs: the existence of a complete set of MUBs implies the decomposition of $M_N(\boldsymbol{C})$ into $N+1$ quasi-orthogonal $N$-dimensional MASAs. Here, we use a term quasi-orthogonal since these subalgebras, as subspaces, are not orthogonal (they have matrices of the form $zI$ in common, where $z$ is non-zero complex number).

So, the conclusion is: as a consequence of mutual unbiasedness of bases $B_j, 1 < j < N+1$, we have a quasi-orthogonal decomposition of $M_N(\boldsymbol{C})$ into $N+1$ $N$-dimensional subspaces (these subspaces are maximal abelian subalgebras by construction) which have matrices of the form $zI$ in common.

The next step is to form a complex projective space from the linear space $M_N(\boldsymbol{C})$. Complex projective space $\boldsymbol{C}P^{N-1}$ can be seen as a set of equivalence classes of the $N$ complex numbers under the relation:

$$(z_0, \dots, z_{N-1}) \sim \lambda(z_0, \dots z_{N-1}),$$

where λ is a non-zero complex number. We denote these equivalence classes by $[z_0, \dots, z_{N-1}]$.

Thus, we can form the complex projective space $\boldsymbol{C}P^{N^2-1}$ from the space $M_N(\boldsymbol{C})$. Moreover, it's linear $N$-dimensional subspaces will become the complex projective spaces $\boldsymbol{C}P^{N-1}$'s under the same map. More details about the geometry of complex projective space and it's role in quantum mechanics can be found in [5].

It was suggested in [9] by Bondal and Zhdanovskiy that a symplectic geometry can be used when approaching the MUB problem.

Complex projective space is a complex manifold which can be equipped with a symplectic structure called the Fubini – Study form. This symplectic structure is exactly what we need to study more closely to be able to prove our theorem.

## 3. Symplectic toric manifolds

In this section, we recall some of the basic results of symplectic geometry. Our focus will be on the symplectic toric manifolds. All results listed here are known and can be found in [6], [7] and [8].

First, we state some of the most important definitions:

Definition 1. A symplectic form $\omega$ on a manifold $M$ is a closed 2-form on $M$ which is non-degenerate at every point of $M$. A symplectic manifold is a pair $(M, \omega)$ where $M$ is a manifold and $\omega$ is a symplectic form on $M$.

Definition 2. A vector field $X$ on $M$ is symplectic if the contraction $i_X\omega$ is closed. A vector field $X$ on $M$ is hamiltonian if the contraction $i_X\omega$ is exact.

Definition 3. A hamiltonian function for a hamiltonian vector field $X$ on $M$ is a smooth function $H : M \to R$ such that $i_X\omega = -dH$.

Definition 4. An action of a Lie group $G$ on a manifold M is a group homomorphism $\psi : G \to$ Diff $(M)$ , where Diff $(M)$ is a group of diffeomorphisms of $(M, \omega)$.

Definition 5. The action $\psi$ is a symplectic action if

$\psi : G \to \mathrm{Symp}(M, \omega) \subseteq \mathrm{Diff}(M)$ , where $\mathrm{Symp}(M, \omega)$ is a group of symplectomorphisms of $(M, \omega)$.

Definition 6. Hamiltonian torus action. If $G = T^n = S^1 \times S^1 \times \ldots \times S^1$ ($n$-dimensional torus), it's action on $M$ is hamiltonian if the corresponding vector field $X_i$ for each component $S^1$ is hamiltonian ie. $i_{X_i}\omega = -dH_i,\ H_i : M \to R$. Putting all these hamiltonian functions together, we get a **moment map μ** $: M \to R^n$.

Definition 7. An action of $G$ is called effective if every element $g \neq e$ from $G$ moves at least one element of $M$.

Definition 8. A symplectic toric manifold $(M^{2n}, \omega, T^n, \mu)$ is a connected and compact manifold $M^{2n}$ of dimension $2n$ carrying a symplectic structure given by the closed and nondegenerate 2-form $\omega$, paired with an effective Hamiltonian action of the standard $n$ – torus $T^n = S^1 \times S^1 \times \dots \times S^1$.

The following theorem is of central importance for our proof.

**Theorem (Atiyah - Guillemin – Sternberg).** Let $(M, \omega)$ be a compact connected symplectic manifold, and let $\psi : T^n \to \mathrm{Symp}(M, \omega)$ be a Hamiltonian torus action with moment map $\mu : M \to \boldsymbol{R}^n$. Then:

a) the levels of $\mu$ are connected,

b) the image of $\mu$ is convex,

c) **the image of $\boldsymbol{\mu}$ is the convex hull of the images of the fixed points of $\boldsymbol{\psi}$**.

This theorem (we will call it AGS theorem in the rest of the paper) is a part of the Delzant's correspondence: there is a bijective correspondence between a symplectic toric manifolds and so called Delzant's polytopes. The bijective map is, actually, the moment map $\mu$.

Now, we want to apply AGS theorem to the complex projective spaces (symplectic form is the Fubini-Study form). Consider the concrete example $(\boldsymbol{C}P^2, \omega_{FS})$ ([6], [8]) with the action of $T^2$ :

$$\left(e^{i\varphi_1}, e^{i\varphi_2}\right)[z_0, z_1, z_2] = \left[z_0, z_1 e^{i\varphi_1}, z_2 e^{i\varphi_2}\right]$$

This action is Hamiltonian and the corresponding moment map is ([6], [8]):

$$\mu([z_0, z_1, z_2]) = \frac{1}{2}\left(\frac{|z_1|^2}{|z|^2}, \frac{|z_2|^2}{|z|^2}\right), \text{ where } |z|^2 = |z_0|^2 + |z_1|^2 + |z_2|^2 .$$

The fixed points of the action are $[1, 0, 0]$ , $[0, 1, 0]$, $[0, 0, 1]$ which are mapped to $(0, 0)$ , $(0, 1/2)$ , $(1/2, 0)$ . So, we obtain a triangle in $\boldsymbol{R}^2$ (the axes are $|z_1|^2/|z|^2, |z_2|^2/|z|^2$). But, we can look at the image of the moment map as an object in $\boldsymbol{R}^3$ with axes $|z_0|^2/|z|^2$ , $|z_1|^2/|z|^2$ , $|z_2|^2/|z|^2$ and the moment map becomes $\mu_1([z_0, z_1, z_2]) = \frac{1}{2}\left(0, \frac{|z_1|^2}{|z|^2}, \frac{|z_2|^2}{|z|^2}\right)$. By changing the coordinates, we get also

$\mu_2([z_0, z_1, z_2]) = \frac{1}{2}\left(\frac{|z_0|^2}{|z|^2}, \frac{|z_1|^2}{|z|^2}, 0\right)$ and

$\mu_3([z_0, z_1, z_2]) = \frac{1}{2}\left(\frac{|z_0|^2}{|z|^2}, 0, \frac{|z_2|^2}{|z|^2}\right)$. Fixed points are mapped to $(0,0,0)$, $(0, 1/2, 0)$, $(0,0,1/2)$, $(1/2, 0, 0)$. We can ignore the point $(0,0,0)$ and form the convex hull of the other points. In this way, we obtain the regular 2-simplex embedded in the space $\boldsymbol{R}^3$. So, the conclusion is that the moment map, corresponding to the diagonal action of $T^2$, acting on the space $\boldsymbol{C}P^2$ gives us the regular 2-simplex in $\boldsymbol{R}^3$.

Analogously, $\boldsymbol{C}P^{N-1}$ will be mapped to the regular $(N-1)$- simplex (the moment map corresponds to the diagonal action of $T^{N-1}$) and similar for $\boldsymbol{C}P^{N^2-1}$.

## 4. Proof of main theorem

**Theorem. The existence of a complete set of mutually unbiased bases in $N$ −dimensional Hilbert space implies the existence of a complete set of mutually orthogonal Latin squares of order $N$.**

Proof. Suppose that the complete set of MUBs exists. This implies the existence of quasi-orthogonal decomposition of $\boldsymbol{C}P^{N^2-1}$ into $N+1$ subspaces $\boldsymbol{C}P^{N-1}$'s (and existence of $A$-basis adapted to that decomposition). Label the coordinates of $\boldsymbol{C}P^{N^2-1}$ as $(z_0, z_1 \dots z_{N^2-1})$ and the non-zero coordinates of the subspaces as $(z_0, z_1 \dots z_{N-1}) \dots (z_0, z_{N^2-N+1} \dots z_{N^2-1})$. Observe that for all complex projective subspaces the coordinate $z_0 \neq 0$. We chose this coordinate to correspond to the unit matrix. The next step is to apply AGS theorem to the space $\boldsymbol{C}P^{N^2-1}$ and all the subspaces $\boldsymbol{C}P^{N-1}$'s. We consider the symplectic toric manifold $\boldsymbol{C}P^{N^2-1}$ with the Hamiltonian action of

$$T^{N^2-1}\colon \left(e^{i\varphi_1}, \dots e^{i\varphi_{N^2-1}}\right)[z_0, \dots, z_{N^2-1}] = \left[z_0, z_1 e^{i\varphi_1}, \dots z_{N^2-1} e^{i\varphi_{N^2-1}}\right],$$

which also induces the actions of $T^{N-1}$'s on the subspaces. Now, we repeat the procedure from the previous section with the moment map:

$$\mu([z_0, \dots, z_{N^2-1}]) = \frac{1}{2}\left(\frac{|z_1|^2}{|z|^2}, \dots, \frac{|z_{N^2-1}|^2}{|z|^2}\right).$$

Therefore, the space $\boldsymbol{C}P^{N^2-1}$ will be mapped to the regular $(N^2-1)$-dimensional simplex in the space $\boldsymbol{R}^{N^2}$ (axes are $|z_0|^2/|z|^2$ ,... $|z_{N^2-1}|^2/|z|^2$) and we will call it an $A$ -simplex. We will also get $N+1$ regular simplices (let's call them $P_1$ -simplices) of dimension $N-1$. $P_1$ - simplices are the images of the restrictions of the moment map to the subspaces $\boldsymbol{C}P^{N-1}$'s. It is important to see that the set of all vertices of $P_1$ -simplices and the set of vertices of $A$ -simplex coincide (as a consequence of the existence of $A$-basis and the same group action on space $\boldsymbol{C}P^{N^2-1}$ and the $N+1$ subspaces $\boldsymbol{C}P^{N-1}$'s). In addition, all $P_1$ - simplices and $A$ simplex will have one vertex in common – point on the axis $|z_0|^2/|z|^2$ (above, we chose $z_0$ coordinate to be common non-zero coordinate of all subspaces). Thus, as a consequence of the existence of a complete set of MUBs, we obtain the regular simplex of dimension $N^2-1$ which can be decomposed into $N+1$ regular simplices of dimension $N-1$ such that all these simplices share one vertex. Moreover, this decomposition should be possible for every choice of the common vertex (we can choose coordinates $z_1$, $z_2$,...$z_{N^2-1}$ to correspond to the identity matrix) and this is the case because a regular simplex is the most symmetric polytope that exists ie. there is no vertex of a regular simplex which can be privileged over the other simplex's vertices. Thus, we can decompose the $A$ simplex into $P_1$ -simplices in $N^2$ possible ways. Now, we use the fact that our quasi-orthogonal subspaces are MASAs ie. that the sets of MUB-projectors are the bases of these subspaces. It is not difficult to see that the moment map will send all MUB-projectors from one commuting subspace to the one point on the $P_1$ – simplex (see relations between MUB-projectors and $A$-matrices from section 2, then invert these equations and apply the moment map). Let's call this point the moment point. Moreover, the moment map will do the same for all commuting subspaces and their MUB-projectors (the only difference is that the different moment points will have different non-zero coordinates for different subspaces). As a consequence of the same type of action of the moment map on the MUB-projectors from different subspaces, we can establish the following **rule**: to the moment point of every $P_1$ – simplex ($N-1$ dimensional simplex) we can assign $(N-2)-$ dimensional sub-simplex of $P$ simplex (see section 1) (MUB-projectors = vertices of $P$ simplex will sit in $P_1$ – simplex) and we can do this each time in the same way ($(N-2)-$ dimensional sub-simplex of $P$ –simplex should be related geometrically to the moment point in the same way on every $P_1$ – simplex). We want to establish such **rule** because we want the MUB-projectors (the vertices of $P$ simplices) to be at some distance from each other ie. we want to introduce the distances between the projectors in order to obtain the complementarity polytope. We can do that without ruining the symplectic structure which gives us the regular simplex and it's decomposition into $P_1$ simplices. Of course, again, due to the symmetry of regular simplex, we do the same for every decomposition of $A$ simplex. The rest of the proof goes by induction over $N$. So, firstly we analyze the case $N = 2$ (the only case that can be visualized). We want to apply the above stated **rule** to this case. $A$ simplex is a tetrahedron and it can be decomposed into three $1 -$ simplices with one common vertex and this decomposition can be done in four different ways. Label the projectors from a complete set of MUBs by $B_1 = \{P_1^1, P_2^1\}$ , $B_2 = \{P_1^2, P_2^2\}$ , $B_3 = \{P_1^3, P_2^3\}$. Now consider one decomposition and apply the moment map to the MUB- projectors: $P_1^1, P_2^1$ will be mapped to the one point (moment point) on one of these three simplices and same is true for other two pairs of projectors and other two 1-simplices. Now, since we want these projectors to be at some distance from each other, we apply above stated **rule**: for each of these three 1-simplices we assign to their's moment points (in this case moment points are positioned at the centers of 1-simplices) one MUB-projector which is mapped to them: for instance $P_1^1$ , $P_1^2$, $P_1^3$ . Then, we consider other three decompositions and do the same (of course, we choose different MUB-projectors to sit on different 1-simplices). Actually, our first choice of three projectors dictates our other choices. As a final result, we get tetrahedron and six MUB-projectors sitting at the centers of it's 1-simplices such that projectors from the same MUB sit at 1-simplices which do not have common vertex. It is not difficult to see that, now, these projectors form the complementarity polytope and that there are four facets of this polytope related in the same way to the four vertices of the tetrahedron which means that tetrahedron = $A -$ simplex can be inscribed into the complementarity polytope for $N = 2$ (of course, we have known this fact from the beginning, but we wanted to demonstrate that our **rule** works in this case as a part of the proof by induction). Now, we formulate the inductive hypothesis: suppose that, after applying the same **rule** for $N$- dimensional case, the MUB-projectors=vertices of $P$ simplices will form the complementarity polytope (complementarity polytope is the convex hull of these projectors) and there will be $N^2$ facets of this polytope related in the same way to the $N^2$ vertices of $A$ simplex. We have seen that this is true for $N = 2$. What remains to be proved is that this is true for $N + 1 -$ case. $N + 1 -$ case can be easily reduced to $N -$case: we just remove one of $N + 2$ commuting subspaces and one MUB-projector from each of the remaining subspaces: what we are left with is the the $N -$ case. In this way, we can reduce $N+1-$ case to $N-$ cases in all possible ways, and since our hypothesis is true for $N-$case it must also be true for $N+1-$ case. So, the conclusion is that for every dimension $N$, the existence of a complete set of MUBs implies the existence of the regular simplex of dimension $N^2-1$ that can be inscribed into complementarity polytope such that the vertices of the simplex sit at the centers of the polytope's facets. This completes the proof.

*Acknowledgements.* I would like to thank dr Saša Dmitrović and dr Ingemar Bengtsson for useful discussions and anonymous referee for useful comments. I would also like to thank dr Tatjana Vuković and dr Saša Dmitrović for their help during my PhD studies and Serbian Ministry of Science, Technological Development and Innovation for financial support. Thanks to dr. Borivoje Dakić for giving me the endorsement for arXiv.

## References

[1] Bengtsson I. and Ericsson Å. , *Mutually unbiased bases and the Complementarity Polytope*, Open. Sys. Information Dyn. **12**, 107 (2005)

[2] Bandyopadhyay S. , Boykin P. O. , Roychowdhury V. , and Vatan F. , *A new proof for the existence of mutually unbiased bases*, Algorithmica **34**, 512 (2002)

[3] Weiner M. , *A gap for the maximum number of mutually unbiased bases*, Proc. Amer. Math. Soc. **141**, 1963 (2013)

[4] Paterek T. , Dakić B. , Brukner Č. , *Mutually unbiased bases, orthogonal Latin squares and hidden variable models*, Physical Review A, **79**, 012109 (2009)

[5] Bengtsson I. and Życzkowski K. , *Geometry of Quantum States*, Cambridge University Press, 2006.

[6] Bühler M. , *Symplectic Toric Manifolds and Delzant's Theorem*, Semester Paper, EHZ Zürich, 2021.

[7] Cannas da Silva, A. , *Lectures on symplectic geometry* vol. 1764, Lecture notes in Mathematics, Springer – Verlag, Berlin 2006.

[8] Cannas da Silva, A. , "*Symplectic toric manifolds", in: Symplectic geometry of integrable Hamiltonian systems* (Barcelona, 2001), Adv. Courses Math. CRM Barcelona, Birkhäuser, Basel, 2003, pp. 85–173

[9] Bondal A. , Zhdanovskiy I. , *Symplectic geometry of unbiasedness and critical points of a potential*, Primitive Forms and Related Subjects-Kavli IPMU 2014, **83,** 1 (2019)

[10] Bondal A. , Zhdanovskiy I. , *Orthogonal pairs and mutually unbiased bases*, Journal of Mathematical Sciences **216,** 23 (2016)